\documentclass[12pt]{iopart}

\usepackage{amssymb}
\usepackage{tikz}
\usepackage{graphicx}
\usepackage[UKenglish]{babel}
\usepackage{cite}
\usepackage[colorlinks=true,linkcolor=blue]{hyperref}

\usepackage{pifont}

\newcommand{\xmark}{\ding{55}}
            
\newcommand{\pfa}{P_{\mathrm{FA}}}
\newcommand{\pfs}{P_{\mathrm{FS}}}
\newcommand{\paf}{P_{\mathrm{AF}}}
\newcommand{\pas}{P_{\mathrm{AS}}}
\newcommand{\psa}{P_{\mathrm{SA}}}
\newcommand{\psf}{P_{\mathrm{SF}}}

\newcommand{\ts}{\bar{T}_\mathrm{S}}
\newcommand{\tw}{\bar{T}_\mathrm{W}}
\newcommand{\ta}{\bar{T}_\mathrm{A}}
\newcommand{\tf}{\bar{T}_\mathrm{F}}
\newcommand{\tj}{\bar{T}_\mathrm{J}}
\newcommand{\tR}{\bar{T}_\mathrm{R}}

\newcommand{\pj}{P_\mathrm{J}}

\begin{document}

\title{Jamming of multiple persistent random walkers in arbitrary spatial dimension}
\author{M. J. Metson, M. R. Evans, and R. A. Blythe}
\address{School of Physics and Astronomy, University of Edinburgh, Peter Guthrie Tait Road,	Edinburgh EH9 3FD}

\date{\today}

\begin{abstract}
We consider the persistent exclusion process in which a set of persistent random walkers interact via hard-core exclusion on a hypercubic lattice in $d$ dimensions. We work within the ballistic regime whereby particles continue to hop in the same direction over many lattice sites before reorienting. In the case of two particles, we find the mean first-passage time to a jammed state where the particles occupy adjacent sites and face each other. This is achieved within an approximation that amounts to embedding the one-dimensional system in a higher-dimensional reservoir. Numerical results demonstrate the validity of this approximation, even for small lattices. The results admit a straightforward generalisation to dilute systems comprising more than two particles. A self-consistency condition on the validity of these results suggest that clusters may form at arbitrarily low densities in the ballistic regime, in contrast to what has been found in the diffusive limit.
\end{abstract}

\maketitle 

\section{Introduction}

The persistent random walker \cite{Taylor1922,Goldstein1951} is currently gaining traction as a fundamental model system in nonequilibrium statistical mechanics. This resurgence in interest stems from the realisation that the persistent random walker belongs to the larger class of \emph{active particles}, which are entities that convert energy into directed motion \cite{Marchetti2013}. Consequently, they break detailed balance at the microscopic scale. In the formulation that we adopt here, the persistent random walker hops some fixed distance in its current direction as a Poisson process at rate $v$, and reorients at rate $\alpha$. It has a variety of applications, including heat transport in turbulent fluids \cite{Taylor1922}, photon transport in thin slabs \cite{Boguna1999} and bacterial motion \cite{Schnitzer1993}. In this latter guise, the persistent random walker is also known as a run-and-tumble particle.

By now, the properties of a \emph{single} persistent random walker are very well established. Early on \cite{Goldstein1951}, it was recognised that in one dimension the probability distribution of the particle's position in continuous space is governed by the telegrapher's equation (i.e., a differential equation that is second order in both space and time). Other exact results in one dimension include mean first-passage times \cite{Angelani2014}, the relaxation spectrum \cite{Mallmin2019} and large-deviation properties \cite{Mallmin2019b}. A variant of the dynamics that further includes a diffusive component has also been solved for the stationary state, relaxation time and first-passage properties \cite{Malakar2018}. In higher dimensions the stationary distribution for a particle confined to a harmonic trap \cite{Malakar2020}, the  probability of remaining in the upper-half plane \cite{Mori2020}, the perimeter of the convex hull \cite{Hartmann2019} and large-deviation statistics \cite{Proesmans2019} have been established.

In this work, our interest is in interactions between \emph{many} persistent random walkers, the consequences of which remain poorly understood from a microscopic viewpoint. We consider specifically the case of hard-core exclusion, which we implement by placing the particles on a hypercubic lattice in $d$ dimensions, and disallow any hops that would lead to two particles occupying the same site. This model (and variants with a softer exclusion constraint) has been studied from a variety of viewpoints \cite{Thompson2011,Soto2014,Slowman2016,Sepulveda2016,Slowman2017,Kourbane2018,Mallmin2019,Partridge2019,Zhang2019} and is sometimes referred to as a \emph{persistent exclusion process}. Macroscopically, it is expected to exhibit a \emph{motility-induced phase separation} \cite{Cates2015}, which arises generically in active particle systems from a feedback between particles accumulating where the propulsion speed is low and this speed decreasing due to crowding from nearby particles.  Under certain restrictions on the nature of the interactions between particles, this macroscopic clustering can be understood through a coarse-graining operation that yields an equilibrium-like free energy functional \cite{Tailleur2008,Thompson2011,Cates2015}. The case of the exclusion interaction described above falls outside this class, thereby rendering important analyses that appeal more directly to interactions at the individual-particle level.

So far, the persistent exclusion process with two particles in one dimension has been solved exactly, both for the stationary distribution \cite{Slowman2016} and the full relaxation spectrum \cite{Mallmin2019}. A solution can also be found when two particles have a diffusive component to their motion \cite{Das2019}, or when particles are stationary for an exponentially-distributed time while reorienting \cite{Slowman2017}. The full many-body dynamics admits an exact hydrodynamic description when a diffusive component of the dynamics dominates the directed motion and the reorientation \cite{Kourbane2018}. In particular it is shown that a homogeneous density field is unstable to phase separation above a critical density that decreases with increasing P\'eclet number (the ratio of the advective and diffusive lengthscales).

Here, instead, we consider the \emph{ballistic} regime, in which particles move deterministically between exponentially-distributed reorientation events. This corresponds to a scaling limit described in \cite{Slowman2016} within which the two-body problem in one dimension could be solved by determining the mean first passage time associated with entering a jammed state in which neither particle can move. In that work the exact solution for the one-dimensional problem was given.

In this work, we develop the first-passage approach to provide a framework in which to study higher-dimensional  many-particle systems. The key concept we introduce is the decomposition of  the configuration space into \emph{channel states} (where particles are colinear and apt to collide) and the remaining \emph{sea state}. The sea state acts as a higher-dimensional reservoir that surrounds the one-dimensional system. This decomposition allows us to compute the jamming probability for a pair of particles. Although this is an approximation expected to be valid only for large lattices, we find that the probability of being in a jammed state is in good agreement with numerical data even in small systems. 

The paper is organised as follows. In Section~\ref{sec:moddef} we define the model. In Section~\ref{sec:twoparticles} we define decomposition of configuration space into sea and channel states and compute the jamming probability for a pair of particles. In Section~\ref{sec:dilute}, we define a dilute limit within which the jamming probability in the many-particle system can be obtained with a straightforward modification to the two-particle result. By analysing a self-consistency criterion for the dilute regime, we obtain evidence suggesting that particles have a propensity to cluster at any nonzero density, as long as the persistence length of the random walk is sufficiently large.

\section{Model definition}
\label{sec:moddef}

\begin{figure}[ht]
    \centering
    \begin{tikzpicture}[xscale=0.8,yscale=0.8]
        \draw[step=1.0,black,thin] (0,0) grid (3,3);
        \filldraw[color=red!60, fill=red!5, very thick](1,1) circle (0.4);
        \draw[thick][->] (0.8,1) -- (1.2,1);
        \filldraw[color=red!60, fill=red!5, very thick](2,2) circle (0.4);
        \draw[thick][->] (2,2.2) -- (2,1.8);
        \draw[thick,color=black!80][->] (2.3,1.6) [out=-50, in=40] to (2.1,1.1);
        \node[below] at (1.5,0) {\small particle 2 hops};
        
        \draw[thick][->] (3.8, 1.5) to (5.2, 1.5);
        
        \draw[step=1.0,black,thin] (6,0) grid (9,3);
        \filldraw[color=red!60, fill=red!5, very thick](7,1) circle (0.4);
        \draw[thick][->] (6.8,1) -- (7.2,1);
        \filldraw[color=red!60, fill=red!5, very thick](8,1) circle (0.4);
        \draw[thick][->] (8,1.2) -- (8,0.8);
        \draw[thick,color=black!80][->] (7.2,1.5) [out=50, in=130] to (7.8,1.5);
        \node[] at (7.5,1.61) {\color{red}\xmark};
        \node[below] at (7.5,0) {\small particle 1 is blocked};
        
        \draw[thick][->] (9.8, 1.5) to (11.2, 1.5);
        
        \draw[step=1.0,black,thin] (12,0) grid (15,3);
        \filldraw[color=red!60, fill=red!5, very thick](13,1) circle (0.4);
        \draw[thick][->] (12.8,1) -- (13.2,1);
        \filldraw[color=red!60, fill=red!5, very thick](14,1) circle (0.4);
        \draw[thick][->] (14.2,1) -- (13.8,1);
        \node[below] at (13.5,0) {\small particle 2 reorients};
    \end{tikzpicture}
    \caption{\label{fig:jam} A possible evolution from an unjammed configuration (left) to a jammed configuration (right) on a $4\times4$ square lattice. Note that the middle configuration is not jammed, because one of the two particles can move without the need for reorientation.} 
\end{figure}
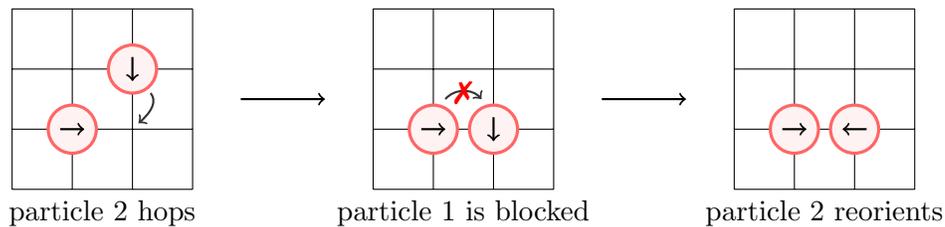

We begin by defining the version of the persistent exclusion process that we study in this work. Particles occupy a periodic $d$-dimensional hypercubic lattice of $L^d$ sites, with the hard-core constraint that no more than one particle can occupy the same site. Associated with each particle is a $d$-dimensional vector $(0,...,0,\pm1,0,...,0)$ that specifies its direction of motion (i.e., along one of the lattice's $d$ principal directions). Each particle hops to the neighbouring site in the direction of motion as a Poisson process with rate $v$, as long as the receiving site is empty (this maintains the hard-core constraint). Without loss of generality, the units of time are chosen so as to set $v=1$. In the literature (e.g.~\cite{Schnitzer1993}), it is conventional for reorientation to be defined as a process that takes place at rate $\alpha$ and causes the particle to adopt one of the $2d$ possible directions of motion with equal probability. This includes the outcome that the particle retains its existing direction of motion. In this work, we find it more convenient to work with the rate
\begin{equation}
	\label{omega}
	\omega = \frac{2d-1}{2d} \alpha \;,
\end{equation}
at which a particle enters a \emph{new} directional configuration. In particular, we find that results in different dimensions are more uniform with this convention. The dynamics of the model are illustrated in Figure~\ref{fig:jam} for the case of two particles and two spatial dimensions.

This model was investigated numerically in \cite{Thompson2011,Soto2014}, solved exactly for two particles in one dimension by \cite{Slowman2016}, and for many particles in a diffusive limit in \cite{Kourbane2018}. To understand what is meant by the latter, let us consider the effect of an additional diffusive contribution to the dynamics, specifically, at rate $D$ a neighbouring site is chosen uniformly and a hop into it attempted. Then let $\bar{x}$ and $\Delta x$ denote the mean and standard deviation in the distance travelled by a particle in the time $\frac{1}{\omega}$ between reorientation events. One finds that $\kappa=\frac{\Delta x}{\bar{x}} = \frac{\sqrt{\omega(D+v)}}{v}$. In Ref.~\cite{Kourbane2018}, a hydrodynamic limit was obtained by taking $D\sim 1$, $v\sim L^{-1}$ and $\omega\sim L^{-2}$. In this limit, the coefficient of variation $\kappa \sim 1$. Here, we consider instead the case where $D=0$ and $\frac{v}{\omega} \sim L$. In this limit, we have $\kappa \sim L^{-1/2}$, and therefore for large $L$ the hopping motion can be considered deterministic while the reorientation is stochastic. We refer to this as a \emph{ballistic} regime. It corresponds to the scaling limit taken in \cite{Slowman2016}.

In the following, we aim to calculate the probability that two particles are in a \emph{jammed} state. This is defined as the situation where the two particles occupy neighbouring sites \emph{and} have opposing velocities.  The right-most configuration shown in Figure~\ref{fig:jam} is an instance of a jammed state: the particles cannot move until one of them reorients. It is this jamming of particles that breaks detailed balance when expressed in its most general form as a kinematic reversibility condition \cite{Mallmin2019}.

\section{Interaction of two particles}
\label{sec:twoparticles}

In this section, our ultimate aim is to estimate the probability that a system comprising only two particles is in a jammed state. This can be found by determining the mean lifetime of a jammed state and the mean time taken to return to a jammed state after it is left. The latter-defines a first passage problem \cite{Redner2001} which we solve approximately by decomposing the configuration space into \emph{sea} and \emph{channel} states. 

\subsection{Decomposition into sea and channel states}
\label{sec:seachannel}

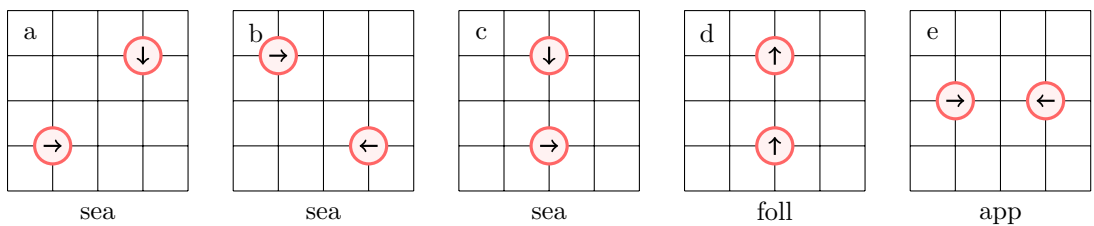
\begin{figure}[ht]
    \centering
    \begin{tikzpicture}[xscale=0.6,yscale=0.6]
        \draw[step=1.0,black,thin] (0,0) grid (4,4);
        \node[] at (0.5,3.5) {\footnotesize a};
        \filldraw[color=red!60, fill=red!5, very thick](1,1) circle (0.4);
        \draw[thick][->] (0.8,1) -- (1.2,1);
        \filldraw[color=red!60, fill=red!5, very thick](3,3) circle (0.4);
        \draw[thick][->] (3,3.2) -- (3,2.8);
        \node[below] at (2,-0.12) {\footnotesize sea};
        
        \draw[step=1.0,black,thin] (5,0) grid (9,4);
        \node[] at (5.5,3.5) {\footnotesize b};
        \filldraw[color=red!60, fill=red!5, very thick](6,3) circle (0.4);
        \draw[thick][->] (5.8,3) -- (6.2,3);
        \filldraw[color=red!60, fill=red!5, very thick](8,1) circle (0.4);
        \draw[thick][->] (8.2,1) -- (7.8,1);
        \node[below] at (7,-0.12) {\footnotesize sea};
        
        \draw[step=1.0,black,thin] (10,0) grid (14,4);
        \node[] at (10.5,3.5) {\footnotesize c};
        \filldraw[color=red!60, fill=red!5, very thick](12,1) circle (0.4);
        \draw[thick][->] (11.8,1) -- (12.2,1);
        \filldraw[color=red!60, fill=red!5, very thick](12,3) circle (0.4);
        \draw[thick][->] (12,3.2) -- (12,2.8);
        \node[below] at (12,-0.12) {\footnotesize sea};
        
        \draw[step=1.0,black,thin] (15,0) grid (19,4);
        \node[] at (15.5,3.5) {\footnotesize d};
        \filldraw[color=red!60, fill=red!5, very thick](17,1) circle (0.4);
        \draw[thick][->] (17,0.8) -- (17,1.2);
        \filldraw[color=red!60, fill=red!5, very thick](17,3) circle (0.4);
        \draw[thick][->] (17,2.8) -- (17,3.2);
        \node[below] at (17,0) {\footnotesize foll};
        
        \draw[step=1.0,black,thin] (20,0) grid (24,4);
        \node[] at (20.5,3.5) {\footnotesize e};
        \filldraw[color=red!60, fill=red!5, very thick](21,2) circle (0.4);
        \draw[thick][->] (20.8, 2) -- (21.2, 2);
        \filldraw[color=red!60, fill=red!5, very thick](23,2) circle (0.4);
        \draw[thick][->] (23.2,2) -- (22.8,2);
        \node[below] at (22,-0.12) {\footnotesize app};
    \end{tikzpicture}
    \caption{\label{fig:seachannel}(a)--(c) Three possible sea configurations; note that although the particles in (c) lie in the same column, they are moving perpendicular to each other, and thus belong to the sea state. (d) A following channel state (foll), in which both particles are moving in the same direction. (e) An approaching channel state (app), in which the particles move in opposing directions.} 
\end{figure}

The notion of sea and channel states arises from the observation that particles are unlikely to jam unless they are both moving along the same line (e.g., a row or column in the two-dimensional case) of the lattice. This line defines a channel, and we require that each particle is moving along the channel. The channel state can be further decomposed into configurations where the particles are both moving in the same direction, and where they are each moving in opposite directions. We call these \emph{following} and \emph{approaching} states, respectively. All other configurations belong to the sea state. Examples of each are given in Fig.~\ref{fig:seachannel}.

The key observations that underpin are analysis are: (i) particles can jam only when they are in a channel state; and (ii) when $L$ is large, entering the channel state from the sea state is a low-probability event, of order $(\frac{1}{L})^{d-1}$, in $d>1$ dimensions. The reasoning behind this second observation is that to enter a channel state, each of $d-1$ coordinates must be the same for both particles, and that the probability of any one coordinate being the same is $\frac{1}{L}$. Note that this estimate relies on the particles having spent sufficiently long in the sea state that their positions are randomised. We can argue this self-consistently, since it implies a sea-state lifetime of order $L^{d-1}$. Simulation results, shown below, further justify this argument.

Note that it is not possible to take $L\to\infty$ at the outset, as was done in the one-dimensional first-passage analysis of~\cite{Slowman2016}, since then the probability of entering a channel vanishes. There are thus two finite length scales in the problem, namely the channel width and the overall system size, which are taken to be $1$ and $L$ respectively.

\begin{figure}
    \centering
    \begin{tikzpicture}[xscale=1,yscale=1]
        \draw[ultra thick][->] (0,0.2) -- (0,1);
        \filldraw[color=red!60, fill=red!5, very thick](0,0.1) circle (0.2);
        \draw[ultra thick][->] (0,2.2) -- (0,3);
        \filldraw[color=red!60, fill=red!5, very thick](0,2.1) circle (0.2);
        \node[] at (0,1.45) {foll};
        
        \draw[->] (1,1.7) -- (2,1.7);
        \node[above] at (1.5,1.7) {$P_{\mathrm{FA}}$};
        \draw[<-] (1,1.3) -- (2,1.3);
        \node[below] at (1.5,1.3) {$P_{\mathrm{AF}}$};
        
        \draw[ultra thick][->] (3,0.2) -- (3,1);
        \filldraw[color=red!60, fill=red!5, very thick](3,0.1) circle (0.2);
        \draw[ultra thick][->] (3,2.8) -- (3,2);
        \filldraw[color=red!60, fill=red!5, very thick](3,2.9) circle (0.2);
        \node[] at (3,1.5) {app};
        
        \draw[->] (4,1.7) -- (5,1.7);
        \node[above] at (4.5,1.7) {$P_{\mathrm{AS}}$};
        \draw[<-] (4,1.3) -- (5,1.3);
        \node[below] at (4.5,1.3) {$P_{\mathrm{SA}}$};
        
        \draw[ultra thick][->] (6,1.2) -- (6,2);
        \filldraw[color=red!60, fill=red!5, very thick](6,1) circle (0.2);
        \draw[ultra thick][->] (6.8,2) -- (7.6,2);
        \filldraw[color=red!60, fill=red!5, very thick](6.7,2) circle (0.2);
        \node[below right] at (6.6,1.5) {sea};
        
        \draw[-] (6,0.5) -- (6,-1) -- (0,-1);
        \draw[->] (0,-1) -- (0,-0.5);
        \node[right] at (6.2,-0.25) {$P_{\mathrm{SF}}$};
        
        \draw[-] (0,3.5) -- (0,4) -- (6,4);
        \draw[->] (6,4) -- (6,2.5);
        \node[right] at (6.2,3.25) {$P_{\mathrm{FS}}$};
    \end{tikzpicture}
    \caption{\label{fig:flow} Probability flow diagram for transitions between the two channel states (following and approaching) and the sea state. The probabilities $P_{\rm XY}$ specify the probability of entering state Y from state X at the points in time when a new state is chosen. Here, the states X and Y are one of A (approaching), F (following) or S (sea). These probabilities are functions of the number of spatial dimensions $d$.}
\end{figure}
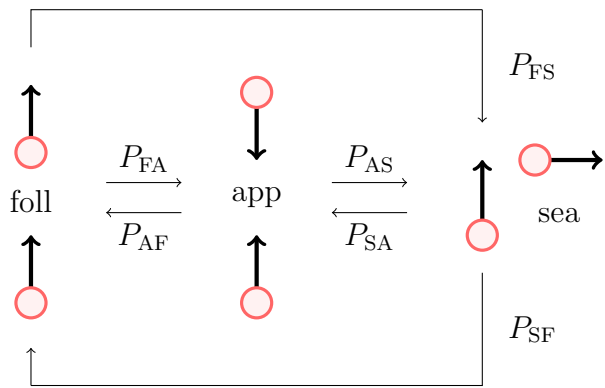

Our aim now is to calculate the mean time to reach a jammed state from other states. Let $\ts$ denote this time starting from the sea state; $\tf(\ell)$ from a separation $\ell$ in the following state; and $\ta(\ell)$ from a separation $\ell$ in the approaching state. Note that the above observations imply that $\ts$ is assumed to be independent of the particle separation within it. Note also that a jammed state is the special case of an approaching state with separation $\ell=0$. To calculate the mean first-passage times, we consider the probability flow between the different states, shown schematically in Figure~\ref{fig:flow}.

Our main interest is in the mean time, $\tR$, for a jammed state to be returned to after a jammed state is left. Exiting a jammed state is achieved by one of the two particles changing its orientation. If the new velocity is opposite to the original velocity, the following state with separation $\ell=0$ is entered. Otherwise, the sea state is entered. Thus, we have that
\begin{equation}
    \label{TRdd}
    \tR = \pas\ts + \paf\tf(0),
\end{equation}
where $\pas$ and $\paf$ are the probabilities of going from the approaching to the sea and to the following states, respectively. These probabilities depend on the number of spatial dimensions (see Section~\ref{sec:pdefs}, below).

In one dimension, there is no sea state: particles are always approaching or following along a single channel. Thus $\pas=0$ and $\paf=1$ and we have that $\tR = \tf(0)$. This case was solved in \cite{Slowman2016} with the result
\begin{equation}
    \tR = \frac{1}{2\omega} + \frac{L}{2} \;.
\end{equation}
This can be understood as follows. The first term is the mean time that the two particles follow each other with zero separation before one of them reverses its direction. Recall that there are two particles, each changing direction at rate $\omega$, so the total rate of velocity reversal is $2\omega$. Once this reversal occurs, there is a probability of $\frac{1}{2}$ that it immediately jams, and of $\frac{1}{2}$ that they enter the approaching state with separation $\ell = L$. The mean time to jam from this state, $\ta(L)$ was shown to be $L$ \cite{Slowman2016}, a result that is curiously independent of the reorientation rate $\omega$. Putting these contributions together, we arrive at the return time given above.

In higher dimensions, we need to determine $\ta(\ell)$, $\tf(\ell)$ and $\ts$. This is achieved by writing down the backward master equations for these quantities, and solving them. The boundary condition on these equations is that $\ta(0)=0$, since the approaching state with zero separation is a jammed state.

The form of these equations is reminiscent of a piecewise deterministic Markov process \cite{Davis1984}, because in the ballistic regime, particle motion is deterministic but reorientation is stochastic. The most straightforward equation to obtain is that for the mean time to jam from separation $\ell$ in the following state. This is
\begin{equation}
    \label{1F}
    \tf(\ell) = \frac{1}{2 \omega} + \pfa \left[ \frac{1}{2} \ta(\ell) + \frac{1}{2}\ta(L-\ell) \right] + \pfs \ts \;.
\end{equation} 
The origin of each term in this expression is as follows. As in the one-dimensional case, the first term is the mean time until one of the two particles adopts a new direction, which in turn causes the following state to be exited. The remaining terms in (\ref{1F}) are a sum over the times to reach a jammed configuration from non-following states, each weighted by the probability that they are entered from the following state. Since the motion is deterministic, the two particles maintain a constant separation $\ell$ in the following state, and then enter the approaching state with separation $\ell$ with probability $\pfa/2$ or with separation $L-\ell$ with probability $\pfa/2$, the latter due to the periodic boundary conditions. Alternatively, with probability $\pfs$, the sea state is entered. Note that as a consequence of (\ref{1F}) we can write the jamming time (\ref{TRdd}) as
\begin{equation}
    \label{TRsimp}
    \tR = \frac{\paf}{2\omega} + \frac{\paf\pfa}{2}\ta(L)  + (\pas + \paf\pfs) \ts  \;,
\end{equation}
due to the boundary condition $\ta(0)=0$.

The backward master equation for the approaching state is more complicated as the separation between particles in this state decreases deterministically at rate $2v$. As in \cite{Slowman2016}, we consider what happens in a short time interval $\delta t$. The system either remains in the approaching state, with smaller separation $\ell - 2 \delta t$ (recall that $v=1$), or enters one of the other two states at some time $\delta t' \le \delta t$ as a result of a reorientation. The probability of the former event is ${\rm e}^{-2\omega \delta t}$, and to account for the latter events we need to integrate over the possible values of $\delta t'$, weighted by the exponential distribution $2\omega {\rm e}^{-2\omega \delta t'}$. We find
\begin{eqnarray}
    \label{1A}
    \ta(\ell) &=& {\rm e}^{-2\omega \delta t} \left[ \delta t + \ta(\ell-2\delta t) \right] \nonumber\\
    && {} + \int_0^{\delta t} {\rm d}{(\delta t')} 2\omega {\rm e}^{-2\omega \delta t'} \left[ \delta t' + \paf\tf(\ell-2 \delta t') + \pas \ts \right] \;.
\end{eqnarray}
We can convert this to a differential equation by expanding to first order in $\delta t$:
\begin{equation}
	\label{1Ad}
	\frac{{\rm d} \ta}{{\rm d} \ell} = \frac{1 + 2\omega [ \paf \tf(\ell) + \pas \ts - \ta(\ell)]}{2} \;.
\end{equation}

The final equation pertains to the sea state. This is relatively straightforward, and reads
\begin{equation}
    \label{1S}
    \ts =  \tw + \int_0^L \frac{{\rm d}{\ell}}{L} \left[ \psa\ta(\ell) + \psf\tf(\ell) \right] \;,
\end{equation}
in which $\tw$ is the average time taken to exit the sea state upon entering it. It is here that the assumption that particles enter the approaching or following states with a separation drawn uniformly on the interval $0 \le \ell \le L$ enters. Strictly speaking, we should sum rather than integrate over the discrete separations $\ell$. This approximation may affect subdominant terms in expressions presented below, but does not, however, affect the emergent scaling behaviour which we discuss in section \ref{sec:dilute}.

\subsection{Solution of the backward master equations}

The system of backward master equations (\ref{1F}), (\ref{1Ad}) and (\ref{1S}) is linear in the first-passage times $\tf(\ell)$, $\ta(\ell)$ and $\ts$. Their solution begins by using (\ref{1F}) to eliminate $\tf(\ell)$ from the other two equations. We do this first for Eq.~(\ref{1S}), obtaining
\begin{equation}
    \label{Ts}
    \ts = \frac{1}{1-\psf\pfs} \left[ \tw + \frac{\psf}{2\omega} + \frac{\psa+\pfa\psf}{L} \int_0^L{\rm d}{\ell}\,\ta(\ell) \right] \;.
\end{equation}
where we have used
\begin{equation}
    \int_0^L {\rm d}{\ell}\, \ta(\ell) = \int_0^L {\rm d}{\ell}\, \ta(L-\ell) \;.
\end{equation}
Our strategy now is to determine $\ta(\ell)$ in terms of $\ts$, since by substituting this into (\ref{Ts}) we can close the system of equations.

To this end, we eliminate $\tf(\ell)$ from Eq.~(\ref{1Ad}) using (\ref{1F}), which yields the differential equation
\begin{equation}
    \label{ODE1}
    \frac{{\rm d}\ta(\ell)}{{\rm d}\ell} = A + B\ta(\ell) + C\ta(L-\ell) \;,
\end{equation}
where
\begin{eqnarray}
	\label{A}
    A &=& \frac{1 + \paf + 2\omega(\pas+\paf\pfs)\ts}{2}\\
    B &=& \omega \left( \frac{\paf\pfa}{2} - 1 \right) \\
    C &=& \omega \left( \frac{\paf\pfa}{2}\right) \;.
\end{eqnarray}
To solve this equation, we make use of the decomposition
\begin{equation}
    \label{1m}
    \ta(\ell) = \underbrace{\frac{\ta(\ell) + \ta(L-\ell)}{2}}_{F(\ell)} + \underbrace{\frac{\ta(\ell) - \ta(L-\ell)}{2}}_{G(\ell)} \;.
\end{equation}
Then, using (\ref{ODE1}), it follows that
\begin{eqnarray}
    \label{1FF}
    \frac{{\rm d}F(\ell)}{{\rm d}\ell} &=& \frac{1}{2} \left[ \frac{{\rm d}\ta(\ell)}{{\rm d}\ell} + \frac{{\rm d}\ta(L-\ell)}{{\rm d}\ell} \right] = (B-C)G(\ell) \\
    \label{1G}
    \frac{{\rm d}G(\ell)}{{\rm d}\ell} &=& \frac{1}{2} \left[ \frac{{\rm d}\ta(\ell)}{{\rm d}\ell} - \frac{{\rm d}\ta(L-\ell)}{{\rm d}\ell} \right] = A + (B+C)F(\ell) \;.
\end{eqnarray}
Differentiating (\ref{1G}), we find $G''(\ell) = (B^2-C^2) G(\ell)$. Noting that, by definition, we must have $G(\ell) = -G(L-\ell)$, the solution of this equation takes the form
\begin{equation}
	G(\ell) = Q \sinh \left( {\textstyle \omega\gamma} [\ell - {\textstyle\frac{L}{2}}] \right)
\end{equation}
where we have a single constant of integration $Q$ and 
\begin{equation}
	\gamma = \sqrt{1-\paf\pfa} \;.
\end{equation}
Substituting this back into (\ref{1G}) we find
\begin{equation}
    F(\ell) =\frac{1}{\gamma} \left[ \frac{A}{\omega\gamma} - Q \cosh \left( {\textstyle \omega\gamma} [\ell - {\textstyle\frac{L}{2}}] \right) \right] \;.
\end{equation}

The constant $Q$ is determined by using the boundary condition $\ta(0)=F(0)+G(0)=0$. This leads ultimately to the expression
\begin{equation}
 \ta(\ell) = \frac{A}{\omega \gamma^2}
\left[ 1 - \frac{\cosh \left( {\textstyle \omega\gamma} [\ell - {\textstyle\frac{L}{2}}] \right) - \gamma \sinh \left( {\textstyle \omega\gamma} [\ell - {\textstyle\frac{L}{2}}] \right) }{ \cosh \left({\textstyle\frac{\omega\gamma L}{2}} \right) + \gamma \sinh \left(  {\textstyle\frac{\omega\gamma L}{2}} \right) } \right] \;.
\end{equation}
Note in particular that $\ta(L)$, which enters into the jamming time (\ref{TRsimp}), takes the relatively simple form
\begin{equation}
	\ta(L) = \frac{2A}{\omega\gamma} \frac{1}{\gamma + \coth \left( {\textstyle\frac{\omega\gamma L}{2}} \right)} \;.
\end{equation}
Combining this with (\ref{A}), we can write the jamming time (\ref{TRsimp}) as
\begin{equation}
\label{TRans}
	\tR = \frac{A}{\omega\gamma} \frac{1 + \gamma \coth\left( {\textstyle\frac{\omega\gamma L}{2}} \right)} { \gamma +  \coth\left( {\textstyle\frac{\omega\gamma L}{2}} \right) } - \frac{1}{2\omega} \;.
\end{equation}

The final part of the solution is to determine the constant $A$. This is achieved by noting that it depends linearly on $\ts$ through Eq.~(\ref{A}), and that $\ts$ depends through (\ref{Ts}) on the integral
\begin{equation}
    \label{kl}
    \frac{1}{L} \int_0^L {\rm d}{\ell}\, \ta(\ell) = \frac{A}{\omega\gamma^2} \underbrace{\left[ 1 - \frac{2}{\omega\gamma L} \frac{1}{ \gamma + \coth\left( {\textstyle\frac{\omega\gamma L}{2}} \right)} \right]}_{\Lambda_L}
\end{equation}
which is proportional to $A$. Then, (\ref{Ts}) and (\ref{A}) imply that
\begin{equation}
\label{Aans}
\fl 2A = \frac{(1 + \paf - \pfs\psf + 2(\pas+\paf\pfs)\omega\tw + \pas\psf)(1-\paf\pfa)}{(1-\paf\pfa)(1-\pfs\psf) - (\pas+\paf\pfs)(\psa+\pfa\psf)\Lambda_L} \;.
\end{equation}
The pair of equations (\ref{TRans}) and (\ref{Aans}) comprise the main result of this paper, namely, the mean time spent between leaving and reentering the jammed state, in terms of a general set of transition probabilities between the channel and sea states.

\subsection{Dimensional dependence of the jamming time}
\label{sec:pdefs}

We now determine how the various transition probabilities that appear in (\ref{Aans}), along with the mean time spent in the sea state $\tw$, depend on the number of spatial dimensions $d$. Recall that we have defined, e.g., $\paf$ as the probability that the system enters the following state (F) after leaving the approaching state (A): see Figure~\ref{fig:flow}. Since the other possibility is to enter the sea state (S), we must have $\paf+\pas=1$. Similarly $\pfa+\pfs=1$ and $\psa+\psf=1$. 

When leaving either the following or approaching state, there are $2d-1$ directions that a particle might choose. Only one of these corresponds to the other channel state. Thus
\begin{equation}
    \label{TP1}
    \paf=\pfa=\frac{1}{2d-1} \qquad \mbox{and} \qquad \pas=\pfs=\frac{2d-2}{2d-1} \;,
\end{equation}
where the latter follows from the conservation of probability. We then also have
\begin{equation}
    \label{TP2}
    \psa=\psf=\frac{1}{2} \;,
\end{equation}
since neither channel state should be favoured upon entry from the sea state. With these definitions, Eq.~(\ref{Aans}) simplifies
considerably to
\begin{equation}
A = \frac{1+ \frac{4(d-1)}{2d-1} \omega \tw}{1 - \Lambda_L} \;.
\end{equation}
Substituting this into (\ref{TRans}) gives
\begin{equation}
\label{trtw}
\fl \tR = \frac{L}{2} \left[ 1 + \frac{4(d-1)}{2d-1} \omega \tw \right] \left[ 1 + \frac{2\sqrt{d(d-1)}}{2d-1} \coth\left( {\textstyle \frac{\sqrt{d(d-1)}}{2d-1}\ \omega L} \right) \right] - \frac{1}{2\omega} 
\end{equation}
where we have used that
\begin{equation}
	\gamma = \sqrt{1-\paf\pfa} = \frac{2\sqrt{d(d-1)}}{2d-1} \;.
\end{equation}

It now remains to determine $\tw$, the mean time spent in the sea state before entering a channel state. The total rate at which a particle changes its direction is $2\omega$. However, when in the sea state, not all such changes cause the channel state to be entered. First, if the two particles initially are parallel or antiparallel to each other, they cannot enter a channel state with a single reorientation. Assuming that all $(2d)^2$ directional configurations are equally likely when in the sea state, we find the probability of not being parallel or antiparallel is $\frac{4d(d-1)}{2d^2}$. Then, of the $2d-1$ new directions that the reorienting particle can adopt, only two lead to it being parallel or antiparallel with the other particle, which is required for a channel state: this event occurs with probability $\frac{2}{2d-1}$. Finally, to be in the channel state, all but one of the particle coordinates must be the same, an event that has probability $(\frac{1}{L})^{d-1}$ as previously noted. Putting these factors together we find that the mean time spent in the sea state before returning to a channel state is
\begin{equation}
    \label{TW}
    \tw = \left( \frac{4d(d-1)}{(2d)^2} \frac{2}{2d-1} \frac{1}{L^{d-1}} 2\omega \right)^{-1} = \frac{(2d-1)dL^{d-1}}{4(d-1)\omega}
\end{equation}
Substituting into (\ref{trtw}), we find
\begin{equation}
	\label{TR}
	\tR = \frac{L}{2} \left( 1 + d L^{d-1} \right) \left[ 1 + \frac{2\sqrt{d(d-1)}}{2d-1} \coth\left( {\textstyle \frac{\sqrt{d(d-1)}}{2d-1}\ \omega L} \right) \right] - \frac{1}{2\omega} 
\end{equation}
The subdominant term in the $(1+dL^{d-1})$ prefactor is crucial to understanding the nature of clustering when the theory is extended to $N$ particles (see section \ref{sec:dilute}), and hence we refrain from dropping it here.

To establish the validity of the assumptions made in the above analysis, we compare the analytical expression for the mean return time (\ref{TR}) with simulation data obtained using a continuous-time Monte Carlo algorithm (see e.g.~\cite{Newman1999,Bortz1975}). Specifically, each particle is assigned a time that the next direction-changing event takes place (that occurs as a Poisson process with rate $\omega$) and, unless it is jammed, a time that the next hop event (rate $v$) occurs. The system then advances to the earliest event in the schedule, and event times recalculated as required. When a reorientation event occurs, each of the $2d-1$ possible new directions is chosen with equal probability. As noted in section \ref{sec:moddef}, we can take $v=1$ without any loss of generality. The independent parameters are then $\frac{\omega}{v}$ and the system size $L$.

To measure the mean return time, we record the current time whenever a jammed state is exited, and subtract this from the time at which a jammed state is next reached. In a system with more than $N=2$ particles, it is possible to immediately reenter a jammed state (i.e., when a third particle is on an adjacent site). In such cases a return time of zero is recorded. We can obtain many (at least $10^4$) return times from a single simulation run in which a jammed state is entered and returned to many times. In the two-particle system, each of these return times is statistically independent: thus the mean return time and the associated statistical error can be determined in the usual way.

\begin{figure}[tb]
    \begin{center}
        \includegraphics[width=0.45\linewidth]{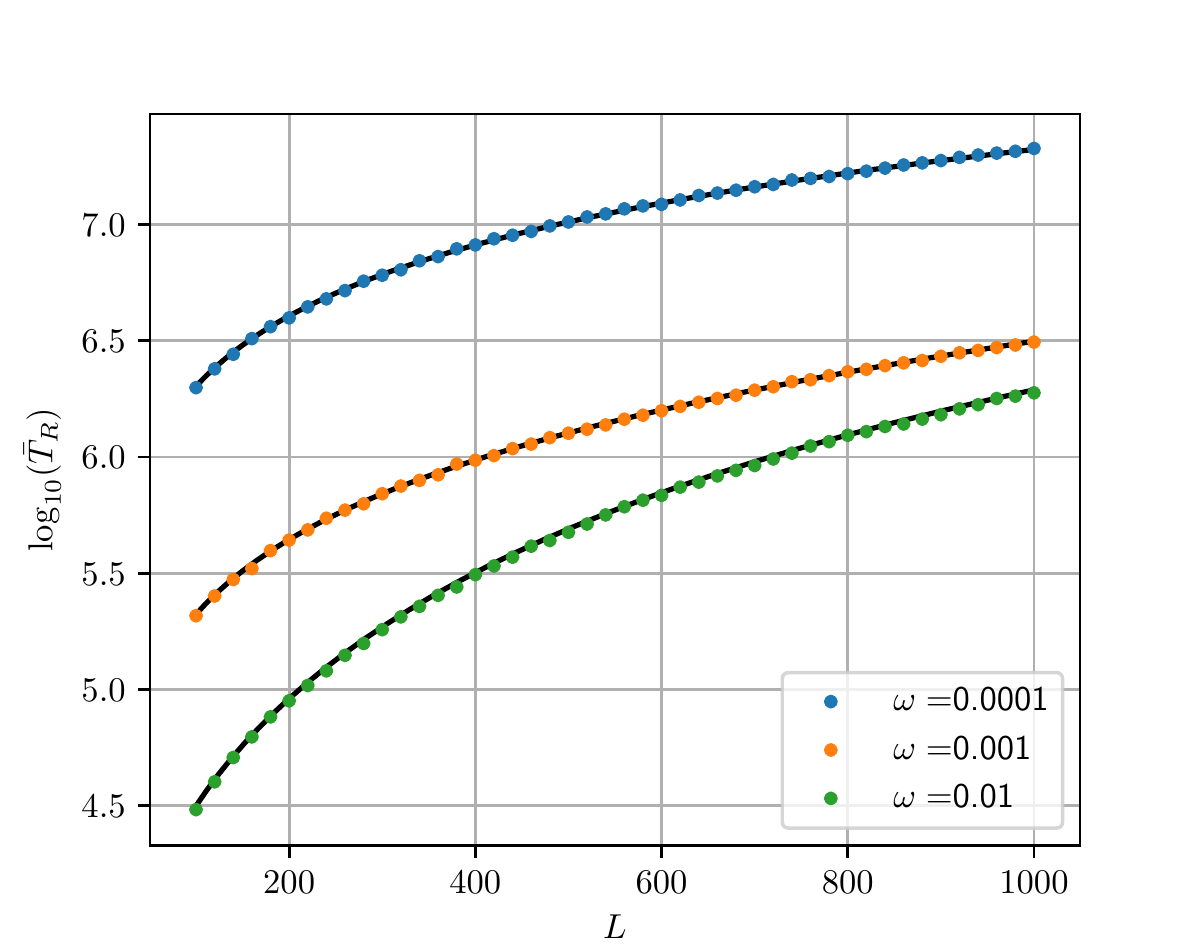}
        \includegraphics[width=0.45\linewidth]{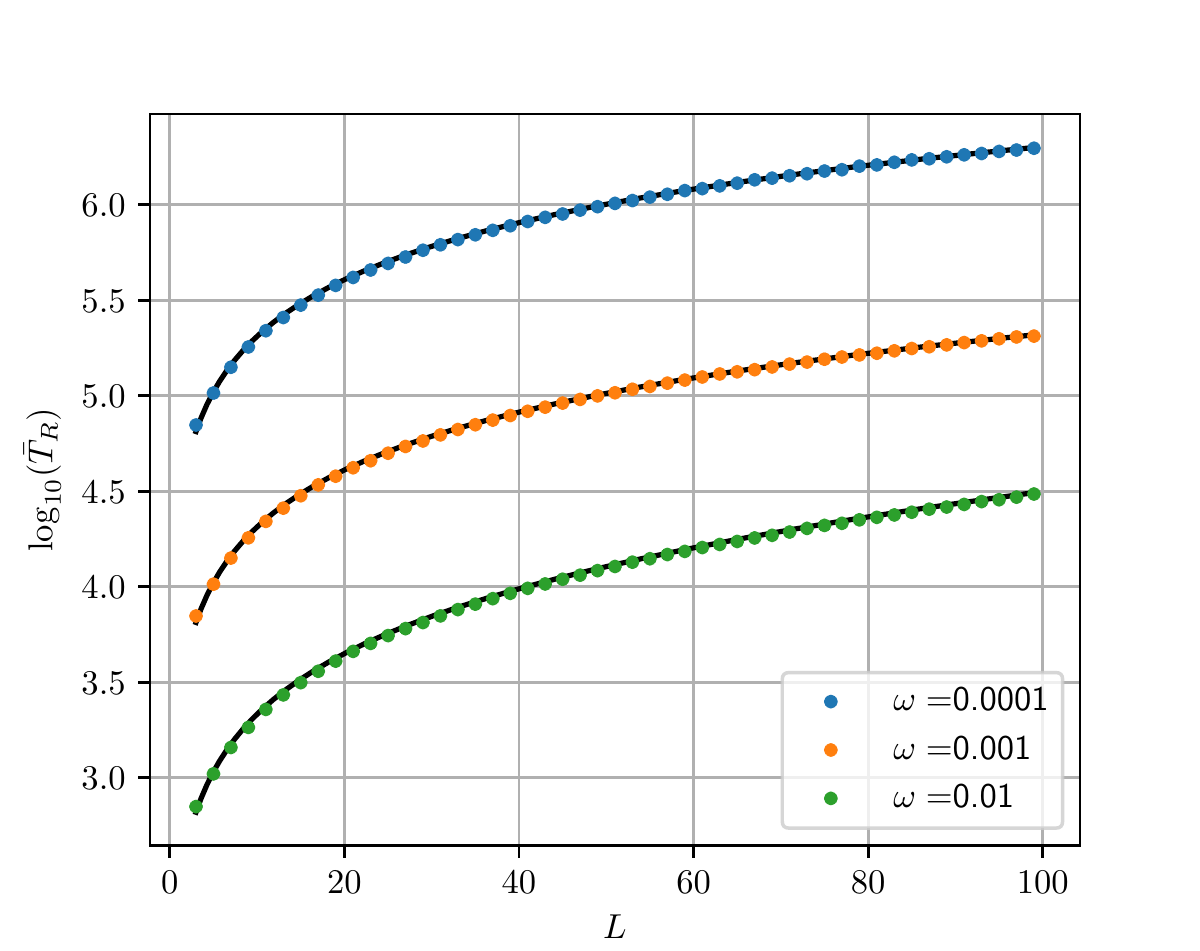}
        \\
        \includegraphics[width=0.45\linewidth]{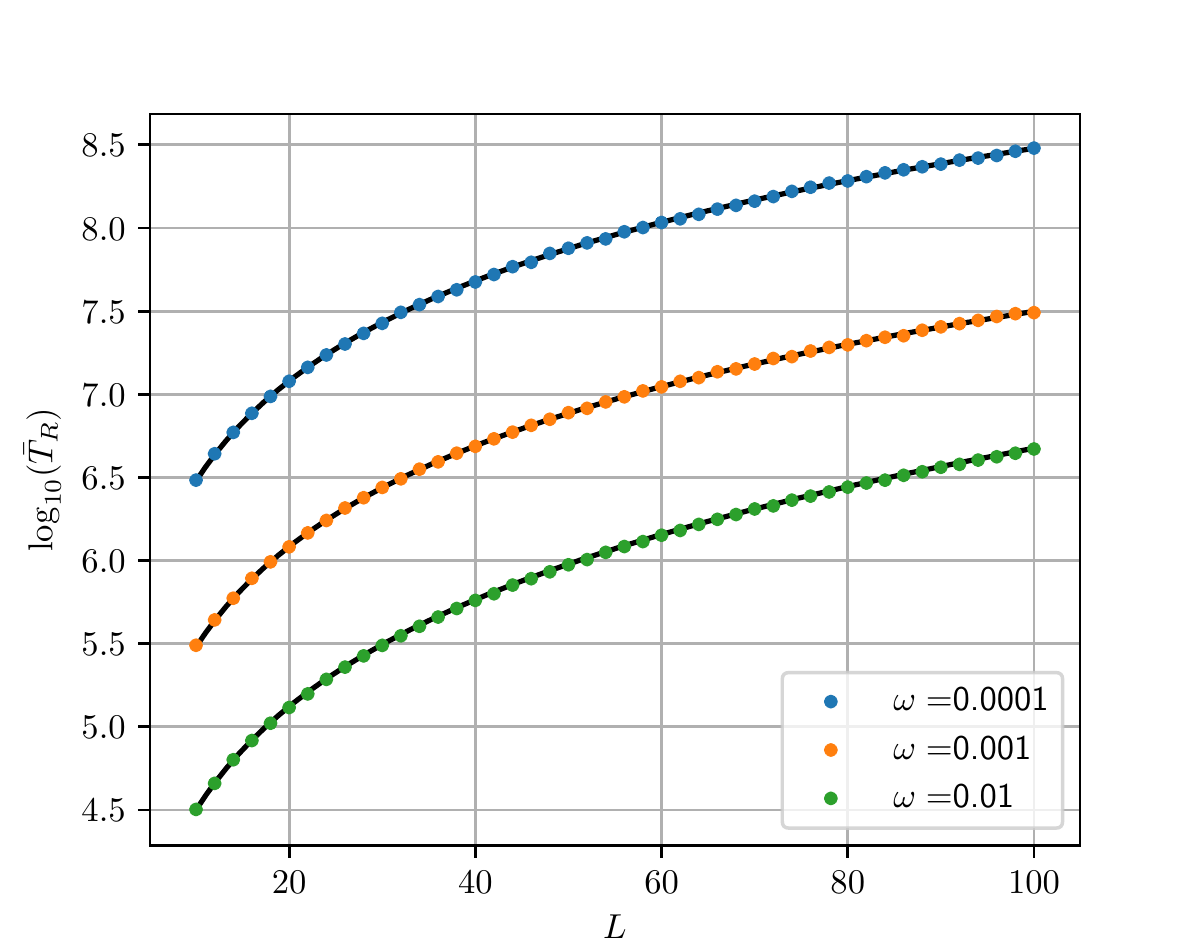}
        \includegraphics[width=0.45\linewidth]{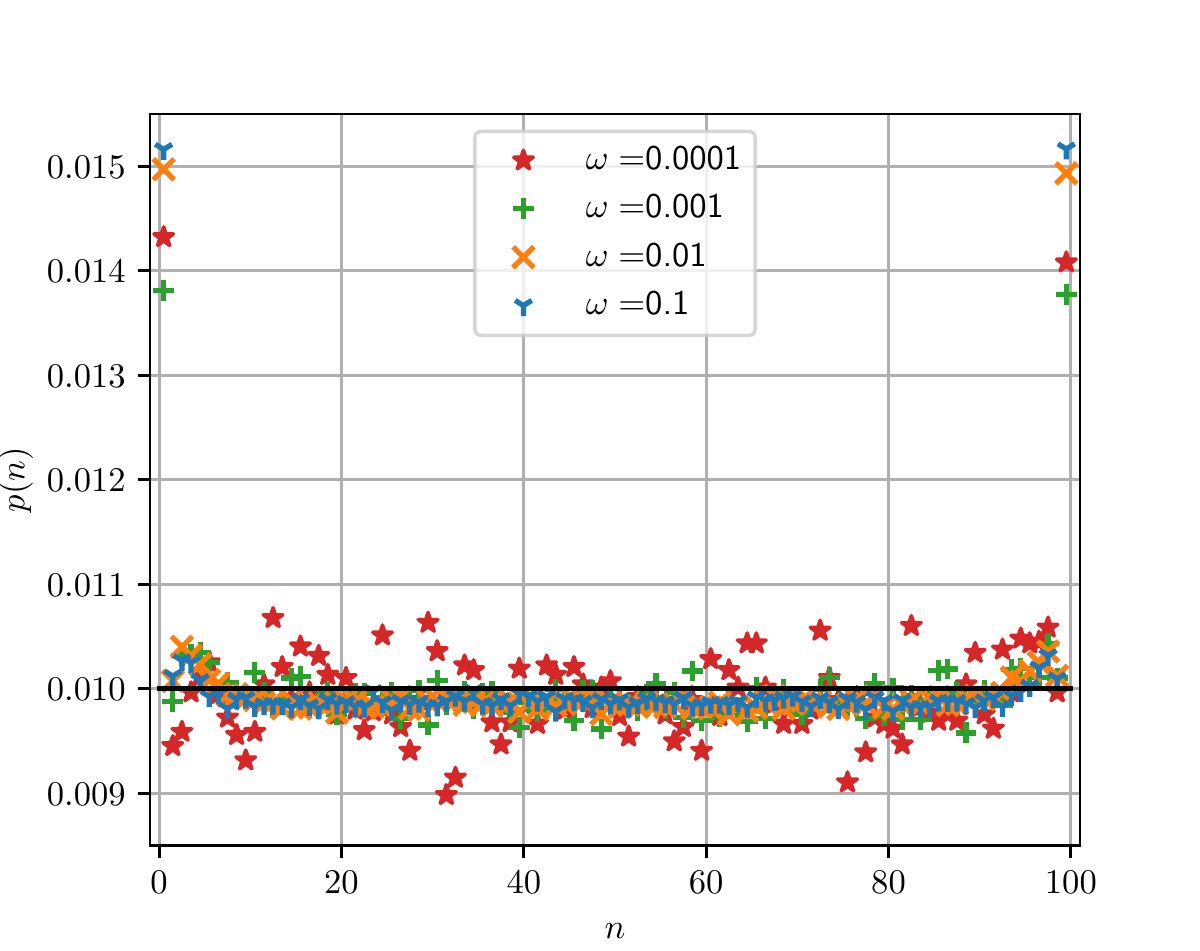}
    \end{center}
    \caption{\label{fig:23d}Upper left, right: logarithmic plots of $\tR$ versus $L$ in 2d for reorientation rates $\omega=10^{-\{2,3,4\}}$ plotted atop (\ref{TR}) (black curves). Deviations from (\ref{TR}) for the $L\gg1$ case (interval $[100,1000]$), shown left, lie in the range $[-2.3,+2.1]\%$ for $\omega=10^{-\{3,4\}}$ bar one anomalous case ($-4.5\%$). For $\omega=10^{-2}$ deviations lie in the range $[-0.3,-4.4]\%$. Excluding $L=3$, the corresponding deviations for smaller $L$ (interval $[3,99]$), shown right, lie in the ranges $[-3.2,+0.3]\%$ and $[-5.1,-1.6]\%$, trending closer to zero in all cases as $L$ increases. Deviations for $L=3$ all lie in the range $[+7.1,+8.2]$. Lower left: corresponding plots for 3d (interval $[10,100]$); deviations in all cases lie in the range $[-2.8,+2.4]\%$. As anticipated, deviations are everywhere largest for $\omega=10^{-2}$. All simulations were run for a minimum of $10^4$ jamming events. Error bars have been omitted since in all cases they are approximately the size of or smaller than the markers. Lower right: distribution of channel-state entry separations, $n$, for the set of 2d systems where $L=101$ and $\omega=10^{-\{1,2,3,4\}}$. A uniform entry distribution corresponds to $p(n)=0.01$ (black line). A minimum of $6\times10^4$ entries were recorded in each case. Increasing noise for decreasing $\omega$ reflects that more simulation time is required to achieve the same number of total entries.}
\end{figure}

The comparison between the analytical and simulation results is shown in Figure~\ref{fig:23d}. In the analysis we assumed both large $L$ and reorientation rates $\omega$ of order $\frac{1}{L}$, as it is then the case that the particle motion between reorientation events can be regarded as deterministic. The figure shows that, in fact, we obtain good agreement with simulation results across a wide range of $L$ and $\omega$. Although the deviation increases as $\omega$ is increased, or as $L$ is decreased, the predictions are within $3\%$ of the numerical values over all simulation conditions shown in Figure~\ref{fig:23d}. This good agreement is found in both two and three dimensions. For larger values of $\omega \ge 10^{-1}$ (not shown), we find that the theory starts to break down with deviations of $10\%$ or more. (We will return to this point below.) We also confirm, in the lower-right panel, that the distribution of particle separations when the channel state is entered is well approximated by a uniform distribution, as in Eq.~(\ref{1S}).

The validity of the approximation even on small lattices ($L<10$) is perhaps surprising. This is likely due to the fact that even here, sufficiently many hopping events can occur between entering and leaving a channel state that the particle distribution is effectively uniform when the channel state is entered.

\subsection{Jamming probabilities}

It is straightforward to convert the return time $\tR$ into the probability $\pj$ of finding the system in a jammed state if we also know $\tj$, the mean time spent in a jammed state. Then, since the system alternates between being jammed and unjammed, we have
\begin{equation}
    \pj = \frac{\tj}{\tR+\tj} \;.
\end{equation}
The mean time spent in the jammed state is straightforward to calculate, since each particle can reorient so as to unjam at rate $\omega$ given by Eq.~(\ref{omega}). Thus $\tj = \frac{1}{2\omega}$ and
\begin{equation}
    \label{dd_jam_prob} \fl
    \pj(d\geq2) = \left\{\omega L(1+dL^{d-1})\left[ 1 + \frac{2\sqrt{d(d-1)}}{2d-1} \coth\left(\frac{\sqrt{d(d-1)}}{2d-1}\ \omega L \right) \right]\right\}^{-1} \;.
\end{equation}
In one dimension and in the continuum limit, the jamming probability has been found exactly as \cite{Slowman2016}
\begin{equation}
    \pj(d=1) = \frac{1}{2\left(1+\frac{\omega L}{2}\right)} \;.
\end{equation}
Here, we find a very similar result in the limit $L\to\infty$ at fixed $\omega L \ll v$. Then, the argument of the $\coth$ function is small while $dL^{d-1}\gg 1$, and from (\ref{dd_jam_prob}) we obtain
\begin{equation}
	\label{pjsimple}
	\pj(d\geq2) \approx \frac{1}{2dL^{d-1}\left(1+\frac{\omega L}{2}\right)} \;.
\end{equation}
We see that the essential difference is the appearance of the prefactor $\frac{1}{dL^{d-1}}$. This is the probability that two randomly-positioned and randomly-oriented particles are travelling within the same channel (either parallel or antiparallel). When $\omega L \ll 1$, it is very likely that a particle traverses the full length of the system before reorienting, and so given that the system is in the channel state, one would expect to see the jammed state with the same probability as in the one-dimensional system. The expression (\ref{dd_jam_prob}) gives an approximation that is valid beyond this quasi-one-dimensional regime. 

We recall that in a system comprising two diffusing particles (i.e., two non-persistent random walkers), each particle configuration is equally likely. Thus, the probability that the two particles are on adjacent sites is $2d/(L^d-1)$. The effect of persistence then is to increase this probability by a factor $L$, at least when the persistence length is long.  At short persistence lengths, we see a crossover to the diffusive $L^{-d}$ scaling, as expected. 

\section{Extension to dilute gases of $N$ particles}
\label{sec:dilute}

Having established our results for two particles we now show how the approach can be extended to an $N>2$-particle system. The two-particle calculation can be extended to $N$ particles in a dilute regime in which it is assumed that the probability of two particles being in a channel is small, and that of more than two particles occupying a single channel can be neglected entirely. We view the dynamics from the perspective of one of the particles in the system. As before, it may be jammed against one of other the particles, in the same channel as another particle, or in the sea. The mean time spent in the jammed state or in the channel is the same as before. However, the mean time spent in the sea, $\tw$, is reduced, because there are now more channels that the particle can exit the sea state into.

More precisely, if the probability of two particles being in a channel is small, then there are in principle $N-1$ channels available, and the probability of entering a channel is increased by a factor of $N-1$. This corresponds to the expression for $\tw$, Eq.~(\ref{TW}), being reduced by the factor $N-1$. Note that we already took into account the probability of the particle in the channel moving in a direction that allows a channel state to be entered, along with the probability that the particle in the sea reorients into the channel. It is now a straightforward matter to substitute the modified expression for $\tw$ into (\ref{trtw}) to obtain the modified return time
\begin{equation}
\label{trN}
\fl	\tR = \frac{L}{2} \Bigg[ \frac{(N-1) + d L^{d-1}}{N-1} \Bigg] \Bigg[ 1 + \frac{2\sqrt{d(d-1)}}{2d-1} \coth\left( {\textstyle \frac{\sqrt{d(d-1)}}{2d-1}\ \omega L} \right) \Bigg] - \frac{1}{2\omega} \;.
\end{equation}
Likewise, the modified jamming probability is
\begin{equation}
	\label{pjN} \fl
    \pj = \left\{ \omega L\Bigg[\frac{(N-1)+dL^{d-1}}{N-1}\Bigg]\Bigg[ 1 + \frac{2\sqrt{d(d-1)}}{2d-1} \coth\left(\frac{\sqrt{d(d-1)}}{2d-1}\ \omega L \right) \Bigg]\right\}^{-1} \;.
\end{equation}
which, when $L\gg 1$ and $\omega L \ll 1$, is approximated by
\begin{equation}
    \pj \approx \frac{N-1}{2dL^{d-1}\left(1+\frac{\omega L}{2}\right)} \;.
\end{equation}
This is a simple factor of $N-1$ larger than the two-particle result (\ref{pjsimple}), which arises from the basic assumption that in the dilute regime, each particle has $N-1$ possible partners when jammed. We note that the term that was subdominant in (\ref{TR}) is of order $N$ here, and therefore may in fact dominate the contribution of order $L^{d-1}$ which was previously dominant. This fact has consequences for cluster formation, as we now describe.

\begin{figure}[tb]
    \begin{center}
        \includegraphics[width=0.45\linewidth]{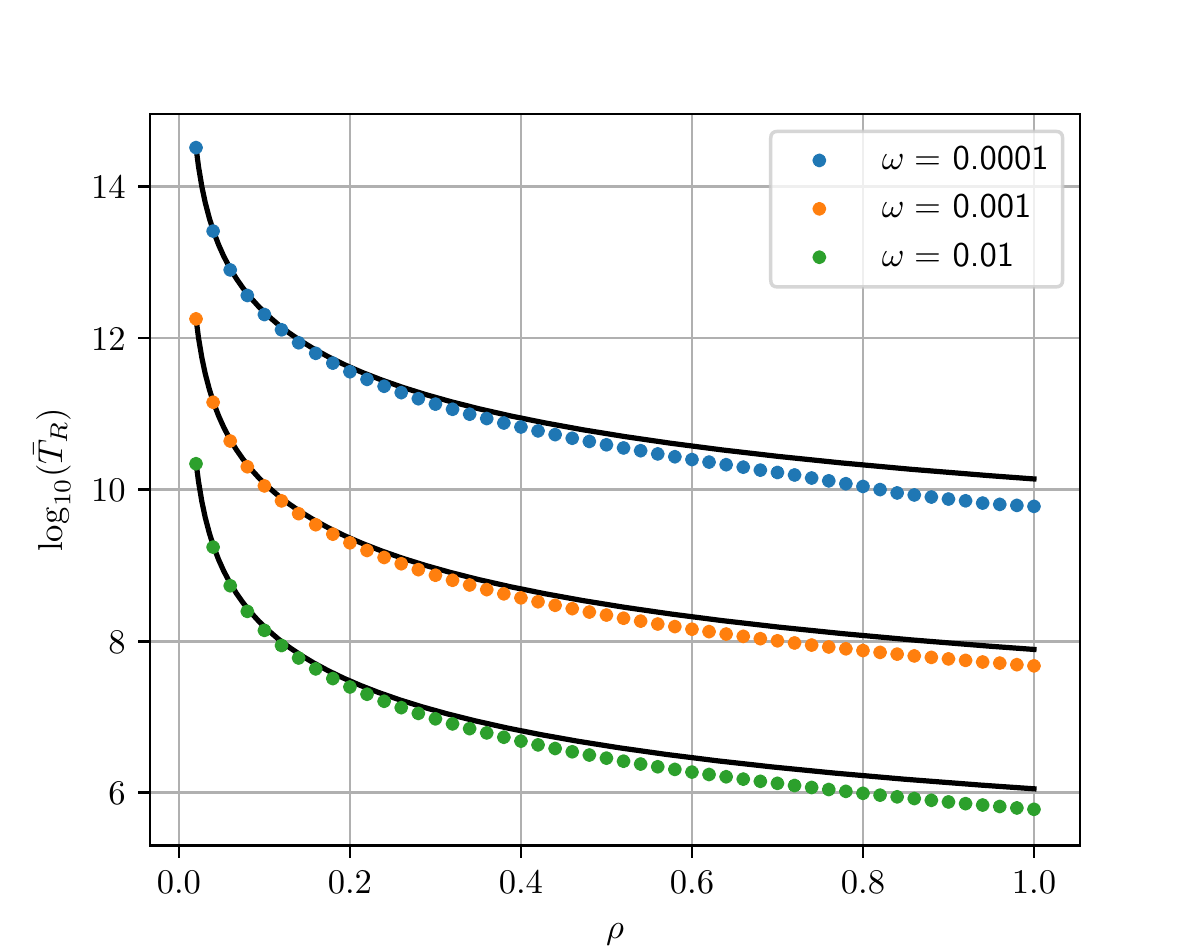}
        \includegraphics[width=0.45\linewidth]{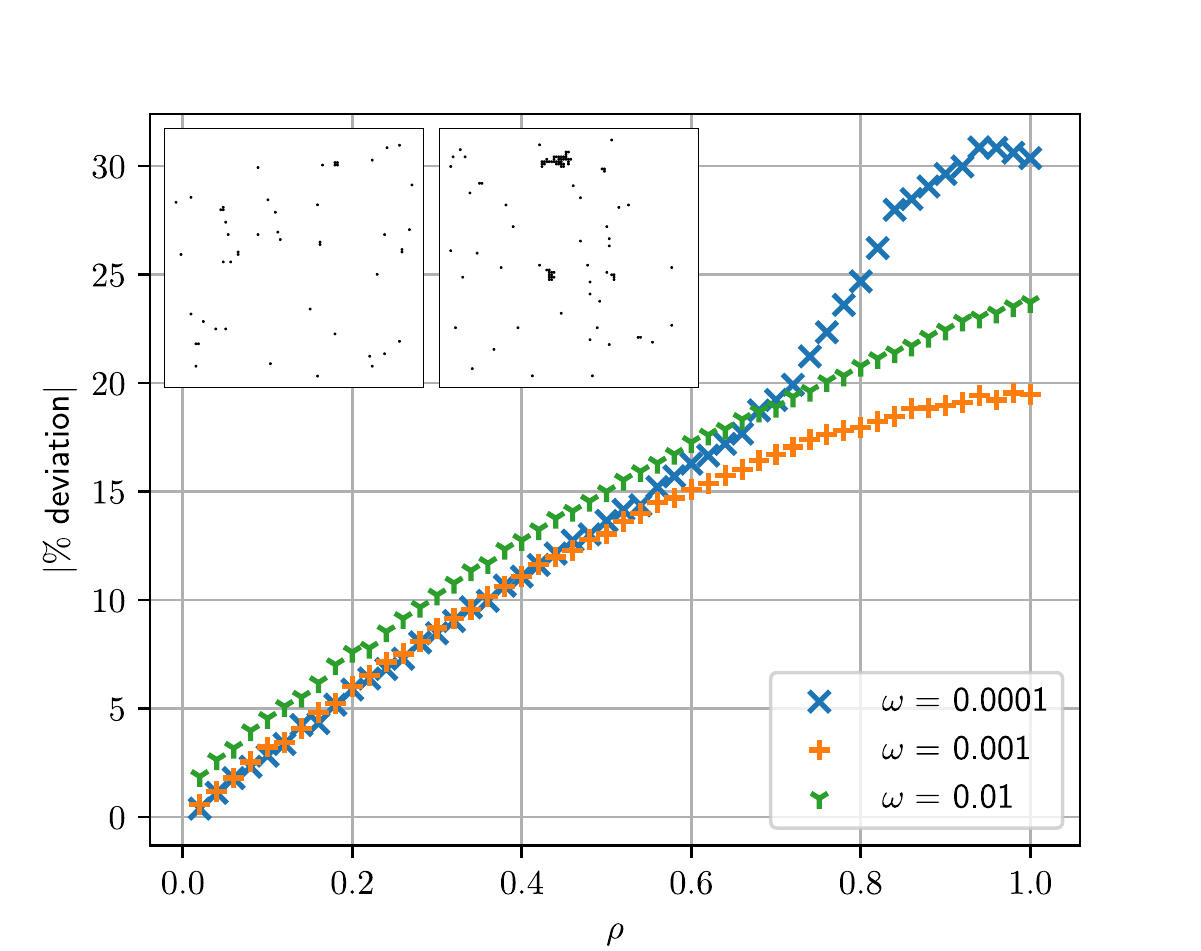}
    \end{center}
    \caption{\label{fig:Nplots} Upper left: simulation results for $\tR$ versus $\rho$ in a 2d $L = 100$ lattice for reorientation rates $\omega=10^{-\{2,3,4\}}$ and $\rho=[0.02,1.00]\%$, plotted atop (\ref{trN}) (black curves). Upper right: deviations from (\ref{trN}). One notes the clustering signature for $\omega=10^{-4}$, evidenced by the system snapshots at $\rho=0.5\%$ and $\rho=1.0\%$ (left and right insets, respectively). Two distinct clusters may be seen in the latter. All simulations were run for $10^6$ jamming events.}
\end{figure}

We are able to gain insight into the formation of clusters because the dilute theory explicitly excludes them. Therefore, deviations of either the return time or the jamming probability from equations (\ref{trN}) or (\ref{pjN}), respectively, are indicative of cluster formation. We demonstrate this in Figure~\ref{fig:Nplots}. At low densities, the simple argument that led to (\ref{trN}) holds reasonably well, with the deviation between this prediction and the return time obtained from simulation slowly increasing as the density increases, as one would expect. Interestingly, however, there is a sudden significant departure of the return time from the predicted value when the reorientation rate $\omega$ is sufficiently small and the density sufficient large. The accompanying simulation snapshots in Figure~\ref{fig:Nplots} suggest that this is due to the onset of particle clustering.
 
To understand this point of departure more deeply, we derive a self-consistency condition for the applicability of the dilute regime. Recall that it was assumed that the probability of more than two particles occupying a channel is negligible. A sufficient condition for two particles to occupy a channel is that they are jammed, so $P_J$ serves as a lower bound on the probability of two or more particles being in the same channel. Therefore, if $P_J$ approaches a value of order $1$, then the theory is no longer self-consistent.

It is revealing to recast (\ref{pjN}) in terms of the particle density $\rho = \frac{N}{L^d}$ and the dimensionless quantity 
\begin{equation}
\xi=\frac{1}{\omega L}\;. 
\end{equation}
We find
\begin{equation}
\label{pjrho}
\pj =  \frac{1}{1+\frac{d}{\rho L}} \frac{\xi}{1+ 2c \coth(\frac{c}{\xi})}
\end{equation}
where $c=\frac{\sqrt{d(d-1)}}{2d-1}$. Viewing this as a function of $\rho$ (with $L$ and $\xi$ both fixed), we see that there is a characteristic density $\rho^* = \frac{d}{L}$, such that when $\rho \ll \rho^*$, the jamming probability increases linearly with $\rho$, and when $\rho \gg \rho^*$ the jamming probability saturates to a value that depends on $\xi$. This characteristic density arises when the dominant and subdominant terms in the jamming probability (\ref{pjN}) are of a similar magnitude.

\begin{figure}[tb]
    \begin{center}
        \includegraphics[width=0.55\linewidth]{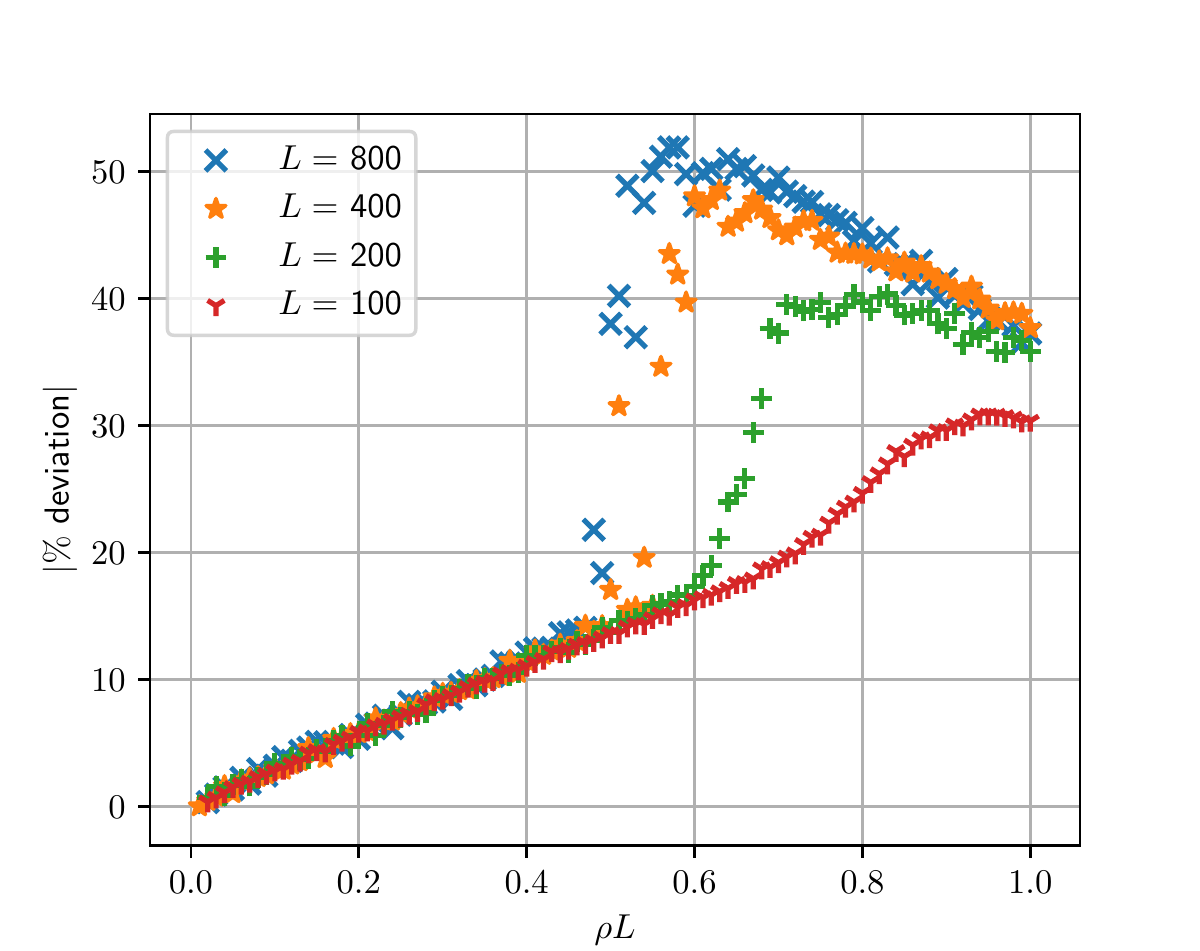}
    \end{center}
    \caption{\label{fig:rho_L_plots}Absolute percentage deviations from (\ref{trN}) in the set of 2d systems where $L=\{100,200,400,800\}$ and $\omega L = 0.01$. All simulations were run for a minimum of $10^5$ jamming events.}
\end{figure}

We recall that due to various approximations that have been made, for example, that hopping is a deterministic process and the replacement of sums with integrals, we may not have captured all the contributions to the subdominant term. Nevertheless, this relation suggests that if we measure the mean return time in different system sizes at fixed $\xi\gg1$ (i.e., fixed $\omega L$), we should expect to see a deviation from the dilute approximation (\ref{trN}) at a value of $\rho$ that scales inversely with $L$. We demonstrate this by plotting the deviation as a function of $\rho L$ as shown in Figure~\ref{fig:rho_L_plots}. We see that, for the case $\omega L =0.01$, there is a significant deviation at $\rho L \approx \frac{1}{2}$, with a sharpening as $L$ increases.  This serves as evidence that clustering occurs at arbitrarily low densities when the persistence length of the random walk is of the order of the system size.

\section{Discussion and outlook}
\label{sec:disco}

In this work we have investigated the persistent exclusion process in more than one spatial dimension. Although the case of two walkers in one dimension is exactly solved \cite{Slowman2016}, the method of solution becomes much more difficult as either the number of particles or the number of dimensions is increased. Consequently, we have developed an approximation scheme that applies in the ballistic regime where the persistence length of the walk is large compared to the lattice spacing, and the lattice is large. This results in our being able to treat hopping as a deterministic process. In this limit, collisions occur only when particles are moving along a common line (which we called a channel) in the higher-dimensional space. By assuming a uniform distribution of particle configurations in the sea state (i.e., when particles do not occupy a single channel), we are able to generalise a method of solution based on first-passage times that is exact in one-dimension \cite{Slowman2016} to higher dimensions. Importantly, this approach may be extended in a very straightforward way to larger numbers of particles.
To this end we defined  a dilute regime,  as one in which at most two particles typically occupy a channel.

The validity of the approximation was checked by comparing with Monte Carlo simulation data for the full stochastic process. For the two-particle system, we find surprisingly good agreement, even on small lattices, as long as the reorientation rate is sufficiently small that fluctuations in the distance moved between direction-changing events can be safely neglected. In particular, we found that a key assumption---that particles are uniformly distributed when they enter a channel state---is well supported by the simulation data. We find that the main effect of persistence in the random walk is for the probability that two particles are jammed to be of order $L$ larger than for diffusing particles when the reorientation rate is sufficiently small. The effective lowering of the dimension by one for persistent random walkers can be understood as a consequence of a collision being very likely when a channel is entered if the persistence length is large.

The multiparticle analysis allowed us to probe the behaviour of systems with a finite particles density, thereby going beyond what has been achieved previously \cite{Slowman2016,Slowman2017,Mallmin2019}. Here, we devised a self-consistency condition for the dilute approximation. The results suggest a characteristic density of order $\frac{d}{L}$ at which the jamming probability begins to saturate. Self-consistency is predicted to break down above this density if the persistence length is comparable with the size of the container. Simulation data are consistent with these predictions: strong deviations  from the dilute approximation, which we interpret as the onset of multiparticle clusters, arise in systems where the particle density exceeds a value of order $\frac{1}{L}$ and the ratio of the hop rate to the reorientation rate is of order $L$. Interestingly, this separation in timescales between the hop and reorientation rates arises naturally in a continuum limit where the lattice spacing is taken to zero \cite{Slowman2016}. This suggests that particles will tend to form clusters at arbitrarily low densities, as long as the persistence length is sufficiently large. 

This result appears to be consistent with the analysis of the persistent exclusion process in the diffusive regime, which shows a decreasing critical density for phase separation as the P\'eclet number is increased and ballistic behaviour is approached \cite{Kourbane2018}. On the other hand, a study of active Brownian particles in the ballistic regime suggests a nonzero critical density for phase separation \cite{Bruss2018}. It is perhaps the case that these systems lack equivalence when persistence lengths are large: in the persistent exclusion process it has been assumed that particles may move only along the principal directions of the lattice, whereas active Brownian particles can move in any direction in continuous space. It would be interesting to generalise the persistent exclusion process to accommodate a larger number of directions of motion to understand if this fundamentally changes its behaviour.

It would also be worthwhile to investigate ways to handle the many-body system beyond the dilute limit, so that the onset of clustering can be understood more deeply. A question that arises in the context of phase separation is whether domains coarsen indefinitely, or become arrested at some finite size. One possible route towards improving on our analysis would be to account for more than two particles in a channel. In particular, if one had a way to characterise the jamming of more than two particles in the one-dimensional system, then a similar embedding of a channel within a sea may provide an adequate description of the higher-dimensional system beyond the dilute regime.

\section{Acknowledgements}
M. J. Metson acknowledges studentship funding from EPSRC through the Scottish CM-CDT under grant number EP/L015110/1.

\bigskip

\bibliography{W1Bib}

\end{document}